\documentclass[twocolumn]{revtex4}
\usepackage{graphicx}% Include figure files
\usepackage{dcolumn}% Align table columns on decimal point
\usepackage{bm}% bold math

\begin{document}

\title{GLASMA EVOLUTION IN PARTONIC MEDIUM}

\author{A.V. Nazarenko}\email{nazarenko@bitp.kiev.ua}
\affiliation{Bogolyubov Institute for Theoretical Physics,\\
14-b, Metrologichna Str., Kiev 03680, Ukraine}

\date{\today}

\begin{abstract}
We examine a scenario of the abelianized Glasma evolution with accounting for
back-reaction of partonic medium in ultrarelativistic heavy-ion collisions.
We announce that such a generalization leads to the instabilities and the
presence of negative color conductivity in the system.
\end{abstract}



\maketitle

%%%%%%%%%%%%%%%%%%%%%%%%%%%%%%%%%%%%%%%%%%%%%%%%%%%%%%%%%%%%%%%%%%%%%%%%%%%%
\section{INTRODUCTION}

Phenomenological analyses of experimental data indicate that the quark-gluon
plasma (QGP) can be formed in ultrarelativistic A+A collisions~\cite{exp}.
Its local thermalization and isotropization should be mainly related to the
fast processes stimulated by instabilities at small times after
collision~\cite{Mrow,Mrow2}.

In the present theoretical picture of ultrarelativistic heavy-ion
collisions~\cite{ETS}, the early stage is preferably characterized by the
large number of partons with ``small'' momenta of the order of the so-called
saturation momentum $\Lambda_s$, which are better viewed as a classical
Yang-Mills field in vacuum~\cite{GV}, sometimes named as ``Glasma''. The initial
conditions for Glasma evolution are determined by the Color Glass
Condensate (CGC) concept by McLerran-Venugopalan (MV)~\cite{McLerran}, where the
field sources {\it before} collision can be presented by the randomly distributed
valent quarks of colliding hadrons and are located (due to Lorentz contraction)
on infinitesimally thin sheets running along the light-cone. These sources are
also treated as the hard partons with ``large'' momenta, which escape quickly
from the system {\it after} collision. Thus, the original MV model neglects the
interaction between the field and the hard partonic medium.

The space-time dynamics of the Yang-Mills fields in vacuum (``the melting of
CGC'') in assumption of boost invariance was investigated numerically, and
the energy and the number distributions of the classically produced gluons
were computed (see, for example, review~\cite{Leon}). Moreover, it was shown
that the violations of boost invariance cause a non-Abelian Weibel
instability~\cite{Wei} leading the field (soft) modes to grow with proper
time~\cite{RV}. However, the effect of isotropization is out of this model.

On the other hand, the hard partons (produced after moment of collision) with
large transverse momentum $p_T$ can be studied within the framework of
transport theory, and if the presence of the soft classical field is neglected,
the time evolution of the these partons is described by Boltzmann equation with a
collision kernel~\cite{GM,BMS,GPZ,MG,DG,NVC} (for comment, see~\footnote{This list
of references contains also some works on approaches based on the relaxation time
approximation to the Boltzmann equation.}). However, it has been argued that the
collective processes caused by the soft gauge field should be dominant in
equilibration of QGP instabilities developed due to anisotropic distributions of
released hard partons~\cite{DN,XG,BMSS,Akk}.

The third regime where the back-reaction of the field on the hard partons
(treated as particles) is still weak but where the self-interaction of the former
may be strongly nonlinear is governed by a ``hard-loop'' effective action which
has been derived in Ref.~\cite{MRS} for arbitrary momentum-space anisotropies. 

It is interesting to note that the numerical studies of anisotropic hard partonic
modes coupled to unstable soft modes revealed the tendency of the non-Abelian gauge
fields to ``abelianize'' during the stage of instabilities~\cite{AL,DN}. It means
that the field commutators become much smaller than the fields themselves.
Moreover, the dynamics of the Abelian and non-Abelian fields is qualitatively the
same, if these fields are not strong enough.

In this paper, we examine a scenario of Glasma evolution with CGC-like initial
conditions, when the presence of the (momentum-)anisotropic medium of hard
partons is also taken into account. Our goal is to evaluate analytically the
behavior of such a system in short-time interval in weak-interaction regime, when
the application of the abelianized version of the field dynamics is possible.
Although the last condition demands to consider the system at relatively large times
after A+A collision (as follows from numerical investigations) but simplifies the
problem considerably. However, it is already pointed out in Refs.~\cite{GKNS,SNK,SKN}
that the early equilibration of QGP is not necessary to describe pion and kaon spectra
observed experimentally at RHIC in Brookhaven.

Since the momentum-space anisotropy of the system can be estimated by means of
transport coefficients, we attempt to calculate a conductivity tensor and to
determine an effect of instabilities on it. It is expected that the back-reaction
can lead to a negative color conductivity in the boost-invariant case.

%%%%%%%%%%%%%%%%%%%%%%%%%%%%%%%%%%%%%%%%%%%%%%%%%%%%%%%%%%%%%%%%%%%%%%%%%%%%
\section{THE MODEL FORMULATION}

As was mentioned in Introduction, the classical Yang-Mills theory in
space-time with pseudo-cylindrical metric 
\begin{eqnarray}
&&ds^2=d\tau^2-\tau^2d\eta^2-dr_T^2-r^2_Td\varphi^2,\\
&&\tau=\sqrt{t^2-z^2},\quad \eta=\frac{1}{2}\ln{\frac{t+z}{t-z}},
\end{eqnarray}
($\tau$ and $\eta$ are proper time and space-time rapidity, respectively) has
been abelianized since $\tau_0\approx3/\Lambda_s$, where
$\Lambda_s\approx2$~GeV \cite{GV}. It means that we actually come to the Maxwell
theory with 4-potential $A_\mu$ (hereafter, we neglect the normalization constant
$1/\sqrt{N_c}$, where $N_c$ is the number of colors).

The free-field theory in mid-rapidity region in the case of central collisions,
when the potentials are parametrized as $A_\tau=0$ (CGC-like gauge fixing),
$A_\eta\equiv\Phi(\tau,r_T)$, $A_{r_T}=0$, $A_\varphi\equiv\Psi(\tau,r_T)$,
has been already examined (see Ref.~\cite{SNK}) in order to describe the space-time
evolution of the field flow (collective velocity) at pre-thermal stage of
collisions. It turns out that the results obtained are qualitatively the same
like in the case of non-Abelian model from Ref.~\cite{KF}.

Here we generalize the abelianized Glasma equations by inclusion of
sources in the right-hand side:
\begin{eqnarray}
&&\partial^2_\tau\Phi-\frac{1}{\tau}\partial_\tau\Phi-\partial^2_{r_T}\Phi-\frac{1}{r_T}
\partial_{r_T}\Phi=J_\eta(\tau,r_T),\\
&&\partial^2_\tau\Psi+\frac{1}{\tau}\partial_\tau\Psi-\partial^2_{r_T}\Psi+\frac{1}{r_T}
\partial_{r_T}\Psi=J_\varphi(\tau,r_T).
\end{eqnarray}
Note that $J_\tau=0$, $J_{r_T}=0$ and the current conservation is satisfied
automatically.

In the context of A+A collisions, the presence of the sources corresponds to
the existence of the medium. Accounting for hard partonic component (viewed as
particle subsystem), we aim to investigate the field and particle dynamics.

The components of the current in Minkowskian space-time are
\begin{equation}
J^\mu=g\int p^\mu(f_+-f_-)\frac{d^3p}{p^0},
\end{equation}
where $p^0\equiv|{\bf p}|$ (the case of massless partons), $f_\pm$ are
distribution functions of (scalar) partons. The distribution of partons is
supposed to be anisotropic in the momentum space and inhomogeneous in
configuration one. Space-time development of functions $f_\pm$ is determined
by Vlasov equations which we will formulate below. Note that ``$-g$'' corresponds
to the charge of electron in the context of electrodynamics.

A toy field model with non-trivial right-hand side has been already investigated
in Ref.~\cite{ALMY}.

It is useful to parametrize momenta as
$(p^\mu)=(p_T\cosh{y},p_T\cos{\phi},p_T\sin{\phi},p_T\sinh{y})$,
where $y$ is momentum rapidity.

In the terms of our variables, one has:
\begin{eqnarray}
&&J_\tau=g\int p^2_T\cosh{\theta}(f_+-f_-)dp_T dy d\phi,\\
&&J_\eta=-\tau g\int p^2_T\sinh{\theta}(f_+-f_-)dp_T dy d\phi,\\
&&J_{r_T}=-g\int p^2_T\cos{\xi}(f_+-f_-)dp_T dy d\phi,\\
&&J_\varphi=-r_T g\int p^2_T\sin{\xi}(f_+-f_-)dp_T dy d\phi,
\end{eqnarray}
where $\theta=y-\eta$, $\xi=\phi-\varphi$.

Taking conditions $J_\tau=0$, $J_{r_T}=0$ into account, the difference
$f_+-f_-$ should be odd function of $\theta$ and $\xi$ during evolution.

The evolution of $f_\pm$ is generated by Vlasov equations:
\begin{equation}
(\hat L\pm g\hat F)f_\pm=0,
\end{equation}
where $\hat L\equiv p^\mu\partial_\mu$, $\hat F\equiv p^\mu F_{\mu\nu}\partial^\nu_p$,
$F_{\mu\nu}=\partial_\mu A_\nu-\partial_\nu A_\mu.$

Since the sources (partons) are randomly distributed at the initial moment
$\tau_0$, we put $f^0_+=f^0_-=f^0$, where $f^0$ is defined as
$(dN_h/d^3xd^3p)|_{\tau_0}$, and $\hat Lf^0=0$ in our investigations. It means that the system
is neutral and the currents are absent at $\tau_0$.

Using the curved coordinates, we obtain
\begin{equation}
\hat L=
p_T\left(\cosh{\theta}\partial_\tau+\frac{\sinh{\theta}}{\tau}\partial_\eta
+\cos{\xi}\partial_{r_T}+\frac{\sin{\xi}}{r_T}\partial_\varphi\right),
\end{equation}
\begin{eqnarray}
&&\hat F=-\frac{\partial_\tau\Phi}{\tau}\partial_y+
\frac{\partial_{r_T}\Phi}{\tau}(\sinh{\theta}\sin{\xi}\partial_\phi
-\cosh{\theta}\cos{\xi}\partial_y)\nonumber\\
&&+\frac{\partial_\tau\Psi}{r_T}(\sinh{\theta}\sin{\xi}\partial_y
+\cosh{\theta}\cos{\xi}\partial_\phi)+\frac{\partial_{r_T}\Psi}{r_T}\partial_\phi.
\end{eqnarray}
We can see that the operator of Lorentz force $\hat F$ is simply the
operator of rotation in momentum space and, therefore, conserves the
absolute value of transverse momentum $p_T$. It is expected that
such structure of the Lorentz force should lead to the momentum
transmission between different directions and, consequently, to the
instabilities in this system.

%%%%%%%%%%%%%%%%%%%%%%%%%%%%%%%%%%%%%%%%%%%%%%%%%%%%%%%%%%%%%%%%%%%%%%%%%%%%
\section{SOLUTION OF EQUATIONS}

In this Section, we concentrate on finding the solution of the set of
the coupled Maxwell--Vlasov equations. Since the Glasma field is essential
at early stage of nuclear collisions (in contrast with the hard partons or
quarks), the study of the field dynamics actually dominates. By this way,
it is necessary to express the particle currents through the fields. In
general, the Vlasov equations are complicated. For this reason, we are
forced to use a method for approximate solving this set.

Fluctuation of distribution function, which arises during fairly small
time interval $\Delta\tau$, can be found in the linear approximation in $g$:
\begin{equation}
f_\pm=f^0\mp g\delta f.
\end{equation}

In this approximation, the space-time evolution of correction $\delta f$,
determining difference $f_+-f_-=-2g\delta f$ and the current components,
results from the following equation:
\begin{equation}
\hat L\delta f=\hat F f^0.
\end{equation}

Note that this approximation does not permit us to investigate an
isotropization of the particle (hard parton) kinematic part of the
energy-momentum tensor, proportional to the sum $f_++f_-$. It is
expected that such isotropization effect can
be observed if the correction of the order $g^2$ is included.
Nevertheless, the approximation under consideration allows one to
study the instabilities in the system.

If $\tau\to\tau_0$, there exists an approximate solution, which is
short-living in time and localized in space,
\begin{equation}\label{appr}
\delta f\approx\frac{\tau-\tau_0}{p_T\cosh{\theta}}\hat F f^0.
\end{equation}

It is easy to verify that the action of evolution operator $\hat L$ on
this expression gives us
\begin{equation}\label{prop}
\hat L\frac{\tau-\tau_0}{p_T\cosh{\theta}}\hat F f^0=\hat F f^0+(\tau-\tau_0)W(\tau),
\end{equation}
where
\begin{eqnarray}
W(\tau)&=&\left[\partial_\tau+\frac{\tanh{\theta}}{\tau}(\partial_\eta+\tanh{\theta})
\right.
\nonumber\\
&+&\left.\frac{\cos{\xi}}{\cosh{\theta}}\partial_{r_T}
+\frac{\sin{\xi}}{r_T\cosh{\theta}}\partial_\varphi\right](\hat F f^0).
\end{eqnarray}

Thus, if $\tau\to\tau_0$, the last term in right-hand side of (\ref{prop})
vanishes. Also note that this is the simplified proof of Eq.~(\ref{appr}).
In order to understand this approximation in details, see Appendix A.

Often the model initial boost-invariant distributions in central heavy-ion collisions
take the form $f^0=f^0(p_T,\theta)=f^0(p_T,-\theta)$ (note that for the sake of
correctness $f^0$ has to be also a function of $r_T$). In this case, we obtain
\begin{equation}\label{Ohm}
J_\eta=\sigma_\eta(\tau)\partial_\tau\Phi, \qquad
J_\varphi=\sigma_\varphi(\tau)\partial_\tau\Psi,
\end{equation}
where
\begin{equation}
\sigma_\eta(\tau)=2(\tau-\tau_0)\sigma_0,\qquad
\sigma_\varphi(\tau)=-(\tau-\tau_0)\sigma_0
\end{equation}
are conductivities.

The common multiplier dependent on the initial distribution of partons is
\begin{equation}
\sigma_0\equiv-2\pi g^2\int\limits_0^\infty dp_T\int\limits_{-\infty}^\infty dy
\partial_y f^0 p_T \tanh{\theta}.
\end{equation}

We can immediately see that $\sigma_\eta(\tau)$ and $\sigma_\varphi(\tau)$
have the different signs. It says about the presence of {\it negative color
conductivity} driving to instability in the system. The mechanism of this
instability looks simple: we deal with situation when the particles (partons)
give the energy to the field.

Now let us analyze the properties of $\sigma_0$. Firstly, we assume that
the initial distribution $f^0$ is the product of the function of $(p_T,\theta)$
and the spatial distribution $(dN_h/d^3x)|_{\tau_0}(r_T)$ in transverse plane.
Taking into account that the initial distribution is even function of $\theta$,
one gets
\begin{equation}
\sigma_0=A\left.\frac{dN_h}{d^3x}\right|_{\tau_0}>0,
\end{equation}
where $A$ is a positive constant arising after integration over momentum
variables.

Thus, $\sigma_\varphi<0$, while $\sigma_\eta>0$. It means that the color
negative conductivity takes place in the transverse plane.

At this stage the natural question arises: how the negative conductivity
does look in the laboratory reference frame. Eqs.~(\ref{Ohm}) are actually the
Ohm law, where $E_\eta\equiv F_{\tau\eta}=\partial_\tau\Phi$,
$E_\varphi\equiv F_{\tau\varphi}=\partial_\tau\Psi$ are the (chromo)electric
field strengths. Introducing $E_i=F_{ti}$ in the Minkowskian space-time,
we find that $E_\eta=\tau E_z$, $E_\varphi=r_T\cosh{\eta}(-\sin{\varphi}E_x+
\cos{\varphi}E_y)+r_T\sinh{\eta}(-\sin{\varphi}F_{zx}+\cos{\varphi}F_{zy})$.
In these terms the current components are
\begin{eqnarray}
J_t&=&-\sinh{\eta}\sigma_\eta E_z,\quad
J_z=\cosh{\eta}\sigma_\eta E_z,\\
J_x&=&-\frac{\sin{\varphi}}{r_T}\sigma_\varphi E_\varphi,\quad
J_y=\frac{\cos{\varphi}}{r_T}\sigma_\varphi E_\varphi.
\end{eqnarray}

If $\eta=0$ and $\varphi=0$, one has that $J_t=J_x=0$, $J_y=\sigma_\varphi E_y$,
$J_z=\sigma_\eta E_z$. Thus the color negative conductivity takes place under 
some conditions (related with the value of angles) in the laboratory reference
frame. This effect is associated with filamentation in the plasma~\cite{Mrow2}.

Since it is hard to find the general solution of field equations for
arbitrary distribution $(dN_h/d^3x)|_{\tau_0}$, we try to study the particular
case, when $(dN_h/d^3x)|_{\tau_0}={\rm const}$. This assumption
simplifies the problem significantly.

When $\sigma_0$ is a constant, the spatial dependence of the field potentials
is immediately derived by using the Bessel-Fourier transform:
\begin{eqnarray}
\Phi(\tau,r_T)&=&\int\limits_0^\infty\Phi_0(k_T)g_\eta(\tau,k_T)J_0(k_Tr_T)dk_T,\\
\Psi(\tau,r_T)&=&r_T\int\limits_0^\infty\Psi_0(k_T)g_\varphi(\tau,k_T)J_1(k_Tr_T)dk_T,
\end{eqnarray}
where the initial conditions resulting from CGC concept are applied:
\begin{eqnarray}
&&g_\eta|_{\tau_0}=0, \quad \left.\frac{\partial_\tau g_\eta}{\tau}\right|_{\tau_0}=k_T,\\
&&g_\varphi|_{\tau_0}=1, \quad \tau\partial_\tau g_\varphi|_{\tau_0}=0.
\end{eqnarray}

In principal, functions $g_\eta(\tau,k_T)$, $g_\varphi(\tau,k_T)$ can be
expressed for arbitrary $\tau_0\not=0$ in terms of Heun functions.
However, these expressions are complicated for heuristic analysis of our model
and its applications.

For this reason, we write down the functions $g_{\eta,\varphi}$ at $\tau_0\to0$:
\begin{widetext}
\begin{equation}
g_\eta=-\frac{k_T}{2\sigma_0}\exp{\left(\frac{1}{2}\sigma_0 \tau^2\right)}
M\left(-\frac{k^2_T}{4\sigma_0},
\frac{1}{2};-\sigma_0\tau^2\right),\qquad
g_\varphi=\frac{1}{\tau}\sqrt{\frac{2}{\sigma_0}}
\exp{\left(-\frac{1}{4}\sigma_0\tau^2\right)}
M\left(\frac{k^2_T-\sigma_0}{2\sigma_0},0;\frac{1}{2}\sigma_0\tau^2\right),
\end{equation}
\end{widetext}
where $M(a,b;z)$ is the Whittaker function.

It is easy to verify that the occurrence of negative color conductivity $\sigma_\varphi$
leads to a growth of the some components of magnetic and electric fields in
comparison with the case of the theory without partonic medium. It is important that the Abelian
magnetic field exhibits a growth, draining some energy from the particle reservoir.
The instabilities are related with the presence of exponents in functions
$g_{\eta,\varphi}$; Whittaker functions change actually a phase of oscillations only
in comparison with the free theory.

If $\sigma_0\to0$ and $\tau_0\to0$, we come to the well-known expressions
(see, for example, Ref.~\cite{Lappi} and references therein):
\begin{equation}
g_\eta(\tau,k_T)=\tau J_1(k_T\tau),\quad
g_\varphi(\tau,k_T)=J_0(k_T\tau),
\end{equation}
where $J_n(z)$ is the Bessel function.

These expressions correspond to the perturbative (lowest order in
the source charge densities) solution.

Now it is necessary to determine the functions $\Phi_0(k_T)$ and $\Psi_0(k_T)$.
They originate from the initial conditions for the field equations.

Note that $\Psi_0(k_T)$ and $\Phi_0(k_T)$ are fluctuating
quantities within the CGC concept, and the pair correlator of the (Yang--Mills)
potentials is observable only. However the field potentials in our approach are not
stochastic quantities in the contrast with CGC ideology because we goal to constitute
the initial conditions on the base of the statistically averaged components of
the energy-momentum tensor accounting for the spatial inhomogeneity.

In order to derive $\Psi_0(k_T)$ and $\Phi_0(k_T)$, let us use the energy
density distribution and the requirement of the absence of field flow at the
initial moment. In mid-rapidity region ($\eta=0$) and $\varphi=0$ (note that
transverse directions are equal in the system with cylindrical symmetry), when
$T_{tt}=T_{\tau\tau}$, $T_{tx}=T_{\tau x}$, we have that
\begin{eqnarray}
T_{tt}|_{\tau_0}&\equiv&\varepsilon(r_T)\nonumber\\
&=&\frac{1}{2}\left(\left.\frac{\partial_{r_T}\Psi}{r_T}\right|_{\tau_0}\right)^2
+\frac{1}{2}\left(\left.\frac{\partial_\tau\Phi}{\tau}\right|_{\tau_0}\right)^2,\\
T_{tx}|_{\tau_0}&=&0,
\end{eqnarray}
where $\varepsilon(r_T)$ is assumed to be the known function from numerical calculations
or physical point of view.

Our trick consists in division of the energy density between different
field components:
\begin{equation}
\partial_{r_T}\Psi|_{\tau_0}=\sqrt{\alpha}r_Tf(r_T),\quad
\left.\frac{\partial_\tau\Phi}{\tau}\right|_{\tau_0}=\sqrt{1-\alpha}f(r_T),
\end{equation}
where $f(r_T)\equiv\sqrt{2\varepsilon(r_T)}$ and $\alpha$ is a some separation
constant (in general, $\alpha$ should be function of $r_T$). Since the potentials
$\Psi$, $\Phi$ are real, one has that $0\leq\alpha\leq1$.

In principal, $\alpha$ is arbitrary constant. Practically, it turns out that
$\alpha\approx1/2$ (it follows from comparison of electric and magnetic strengths
within the numerical approach). Note that the observables of the source-free
theory are independent on $\alpha$.

Thus, one finds
\begin{equation}
\Psi_0(k_T)=\sqrt{\alpha}\tilde f(k_T),\quad
\Phi_0(k_T)=\sqrt{1-\alpha}\tilde f(k_T),
\end{equation}
here
\begin{equation}
\tilde f(k_T)=\int\limits_0^\infty f(r_T)J_0(k_Tr_T)r_Tdr_T.
\end{equation}

These expressions finally determines the fields in our model.

%%%%%%%%%%%%%%%%%%%%%%%%%%%%%%%%%%%%%%%%%%%%%%%%%%%%%%%%%%%%%%%%%%%%%%%%%%%%
\section{APPLICATIONS}

In the previous Sections we have formulated the model of Glasma in hard partonic medium.
Since the classical field modes are usually interpreted as soft partons, their spatial dependence
at early stage of A+A collision may be done within the framework of Glauber model. 
However, the explanations of experimental data can be efficiently done with application of
another distributions too. As it was demonstrated in Ref.~\cite{SKN}, the Gaussian
distribution of soft partons leads to the adequate pion spectra produced after collision
at RHIC. To formulate the field initial conditions, here we would like to choose the same
approximation for the energy density at the initial moment,
\begin{equation}
\varepsilon(r_T)=E\exp{\left(-\frac{r^2_T}{2R^2}\right)},
\end{equation}
where $E=45$~GeV/fm${}^3$, $R=3.768$~fm.

Then we find that
\begin{equation}
\tilde f(k_T)=2^{3/2}\sqrt{E}R^2\exp{(-k_T^2R^2)}.
\end{equation}

The boost-invariant distribution function $f^0$ (defining conductivities) is
completely arbitrary at this point, so in order to proceed one needs to assume a
specific form for it. In what follows we will require that $f^0$ is obtained
from isotropic function,
\begin{equation}\label{dist0}
N_0\exp{\left(-\frac{p^0}{p_h}\right)},
\end{equation}
by the replacement $y\to\theta$ in $p^0=p_T\cosh{y}$ and by the rescaling of
one dimension in momentum space,
\begin{equation}\label{dist1}
f^0=N(\zeta)\exp{\left(-\frac{p_T}{p_h}\sqrt{\cosh^2{\theta}+\zeta\sinh^2{\theta}}\right)},
\end{equation}
where $p_h$ takes the role of saturation moment, $\zeta>-1$ is a parameter reflecting
the strength of the partonic medium anisotropy and $N(\zeta)$ is a normalization constant.
Note that $\zeta>0$ corresponds to a contraction of the distribution in the $z$-direction
whereas $-1<\zeta<0$ corresponds to a stretching of the distribution in the $z$-direction.

Constant $N(\zeta)$ is simply determined by requiring the number density to
be the same both for isotropic and anisotropic systems and can be evaluated
(by integration over momentum variables) to be
\begin{equation}
N(\xi)=N_0\sqrt{1+\zeta}.
\end{equation}

Integrating over momentum, the multiplier defining the conductivities is
\begin{equation}
\sigma_0=4\pi g^2 N_0 p^2_hC(\zeta),
\end{equation}
where
\begin{equation}
C(\zeta)=\frac{2}{3}(1+\zeta)^{3/2}
F\left(\left[2,\frac{3}{2}\right],\left[\frac{5}{2}\right],-\zeta\right).
\end{equation}

The coefficient $C(\zeta)$ in the region $\zeta\in(-1,\infty)$ is determined by
hypergeometric function $F$ and is such that $C(-1)=0$ (the case of source-free
theory), $C(0)=2/3$ (for isotropic medium), $C(\infty)=\pi/2$.

\begin{figure}[htbp]
\begin{picture}(60,45)
\put(-90,-121){\includegraphics[width=8.cm]{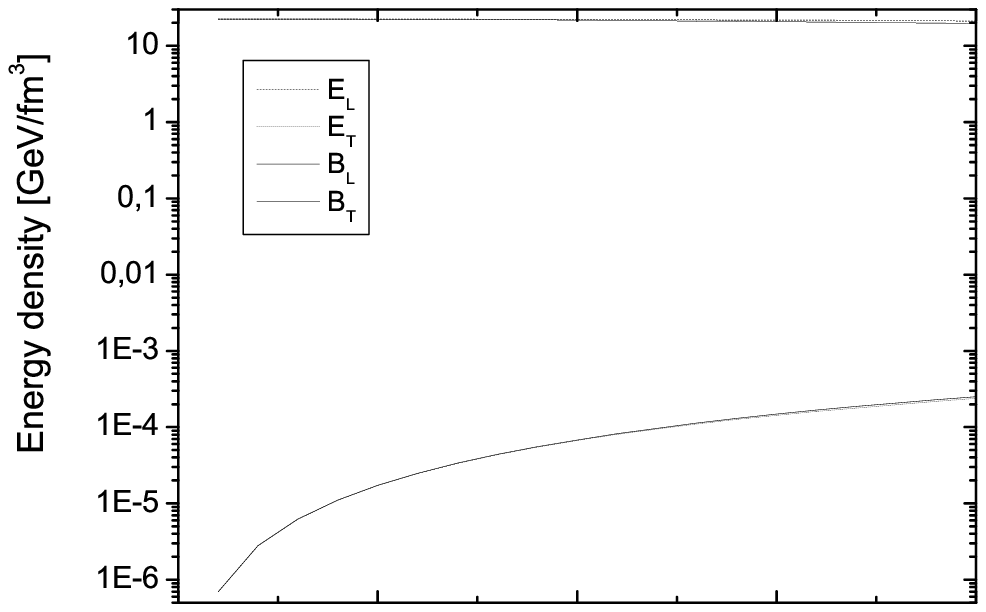}}
\put(-94,-270){\includegraphics[width=8.32cm]{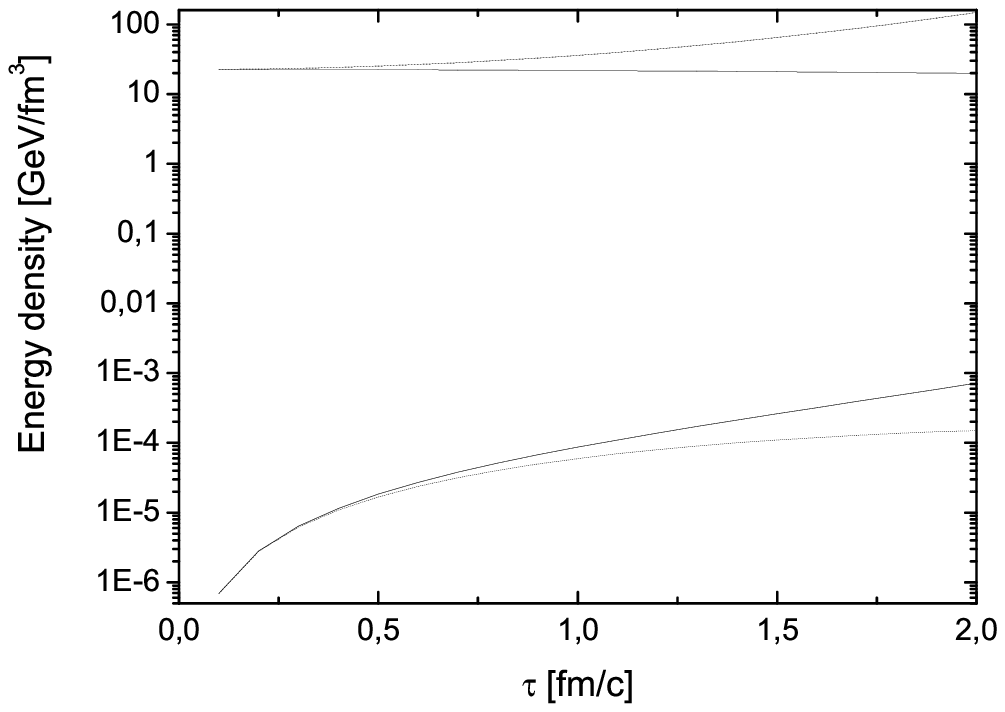}}
\end{picture}
\vspace*{90mm}
\caption{\label{fig} Time evolution of the field energy density split into
longitudinal and transverse electric ($E$) and magnetic ($B$) components at
$r_T=0.1$~[fm], $\eta=0$, $\varphi=0$. The top panel corresponds to the free
theory with $\sigma_0=0$. The bottom panel demonstrates the growth of $E_L$
and $B_T$ at $\sigma_0=0.25$~[fm${}^{-2}$].}
\end{figure}

Fig.~1 shows how the exponentially growing energy transferred from hard to
soft partons is distributed among magnetic and electric fields at
$\sigma_0\not=0$ in comparison with the case of free field theory. The
dominant contribution is still in longitudinal electric field (in accordance
with CGC-like initial conditions). Nevertheless, we see that the transverse
magnetic field demonstrates unstable behavior too while this effect is absent
in the free theory. Since the particle subsystem gives the energy to
the field, the total field energy density (as the sum of components) tends to
grow with time.

Note that a similar model with expanding Abelian field coupled to the hard
partons, when the strict boost invariance of fields is relaxed, has been already
developed in Ref.~\cite{RR} in the context of the quark-gluon plasma.

\section{CONCLUSIONS}
%%%%%%%%%%%%%%%%%%%%%%%%%%%%%%%%%%%%%%%%%%%%%%%%%%%%%%%%%%%%%%%%%%%%%%%%%%%%

Generalizing the space-time evolution of the expanding Glasma with CGC-like
initial conditions by inclusion of small density of the hard partons
anisotropically distributed in momentum space, we observe the instabilities
(due to transferring energy from hard to soft partons) in the case of the
abelianized boost-invariant model. 

Here we propose to measure an anisotropy by means of transport
coefficients like the conductivity tensor in the contrast with the usual
approach based on the energy-momentum tensor. As the result, the
instabilities in the system under consideration lead to a conclusion of the
presence of negative color conductivity in A+A collisions. The sign of
conductivity in transverse plane depends on the angle what says about
filamentation inherent to Weibel instabilities in plasma.

Unfortunately, the approximate solution derived to the transport equations does
not permit us to achieve the isotropization here. This problem should be
investigated in details and will be published elsewhere.

\section*{ACKNOWLEDGEMENTS}
%%%%%%%%%%%%%%%%%%%%%%%%%%%%%%%%%%%%%%%%%%%%%%%%%%%%%%%%%%%%%%%%%%%%%%%%%%%%

The research of author was partially supported by the Foundation of Department
of Physics and Astronomy of NAS of Ukraine.

\appendix
%%%%%%%%%%%%%%%%%%%%%%%%%%%%%%%%%%%%%%%%%%%%%%%%%%%%%%%%%%%%%%%%%%%%%%%%%%%%
\section*{APPENDIX A. PROPAGATOR OF TRANSPORT EQUATION}

Here, we would like to discuss in details the approximation which we have
applied previously. To find $\delta f$, we have to determine operator $\hat L^{-1}$,
inverse to the first-order evolution operator $\hat L$. The inversion procedure of
the evolution operator of transport equation was elaborated by Landau and, generally
speaking, results in emergence of the Landau damping in plasma.

Let $G$ is the solution of the following equation:
\begin{equation}
\hat LG(\tau,\theta,{\bf r}_T|\tau^\prime,\theta^\prime,{\bf r}_T^\prime)
=\frac{\delta(\tau-\tau^\prime)}{\tau}\delta(\theta-\theta^\prime)
\delta^2({\bf r}_T-{\bf r}_T^\prime),
\end{equation}
where right hand side is the 4-dimensional $\delta$-function with respect to the
pseudo-cylindric measure $\tau d\tau d\theta d^2r_T$.

Without loosing generality, we limit ourselves by the case, when $\tau_0\to0$,
and by using the transverse coordinates ${\bf r}_T$ instead of cylindrical
$(r_T,\varphi)$. 

Using the formulas from Appendix B, $G$ is represented as
\begin{widetext}
\begin{equation}
G(\tau,\theta,{\bf r}_T|\tau^\prime,\theta^\prime,{\bf r}_T^\prime)=\lim_{\varepsilon\to+0}
\frac{i}{(2\pi)^4p_T}\int\frac{{\rm e}^{i\omega[\tau\cosh{(\xi-\theta)}-\tau^\prime\cosh{(\xi-\theta^\prime)}]
-i{\bf k}_T({\bf r}_T-{\bf r}^\prime_T)}}
{{\bf k}_T{\bf v}_T-\omega\cosh{\xi}-i\varepsilon}
\omega d\omega d\xi d^2k_T,
\end{equation}
\end{widetext}
where ${\bf v}_T\equiv{\bf p}_T/p_T$ and the Landau damping is already taken into
account due to auxiliary formula:
\begin{equation}
\lim_{\varepsilon\to+0}\frac{1}{x-i\varepsilon}={\cal P}\frac{1}{x}+i\pi\delta(x).
\end{equation}

Discarding the spatial dispersion, we have to assume that $k_T\ll\omega\cosh\xi$.
It leads to simplification:
\begin{eqnarray}
&&G(\tau,\theta,{\bf r}_T|\tau^\prime,\theta^\prime,{\bf r}_T^\prime)\approx
\frac{1}{p_T}\Theta(\tau\cosh{\theta}-\tau^\prime\cosh{\theta^\prime})
\nonumber\\
&&\times\delta(\tau\sinh{\theta}-\tau^\prime\sinh{\theta^\prime})
\delta^2({\bf r}_T-{\bf r}^\prime_T).
\end{eqnarray}
where $\Theta$ is Heaviside function defined as $\Theta(x<0)=0$,
$\Theta(x=0)=1/2$, $\Theta(x>0)=1$.

This approximation says that the system is homogeneous at significantly
large times and the fluctuations are localized in space.

Furthermore, let the Bj\"orken scaling flow, when $\theta\approx0$, take
place. It means that $\theta$ and $\theta^\prime$ should be equal and gives
us that
\begin{equation}
G(\tau,\theta,{\bf r}_T|\tau^\prime,\theta^\prime,{\bf r}_T^\prime)\approx
\frac{\Theta(\tau-\tau^\prime)}{p_T\tau^\prime\cosh{\theta^\prime}}
\delta(\theta-\theta^\prime)\delta^2({\bf r}_T-{\bf r}^\prime_T).
\end{equation}

More precisely, it can be derived from condition, $\tau\sinh\theta={\rm const}$,
resulting in
\begin{equation}
\sinh\theta d\tau+\tau\cosh d\theta=0,
\end{equation}
where $d\tau=\tau-\tau^\prime$, $d\theta=\theta-\theta^\prime$.

Assuming that $\tau$ is small and the expression under integration is not
essentially changed at this time range, we can do the following replacement:
\begin{equation}
\int\limits_0^\infty d\tau^\prime\Theta(\tau-\tau^\prime)F(\tau^\prime)
\to \tau F(\tau).
\end{equation}

Then, one obtains that
\begin{eqnarray}
&&\int G(\tau,\theta,{\bf r}_T|\tau^\prime,\theta^\prime,{\bf r}_T^\prime)
F(\tau^\prime,\theta^\prime,{\bf r}_T^\prime)
\tau^\prime d\tau^\prime d\theta^\prime d^2r^\prime_T\approx\nonumber\\
&&\approx\frac{\tau}{p_T\cosh{\theta}}F(\tau,\theta,{\bf r}_T).
\end{eqnarray}
This formula determines the solution of inhomogeneous transport equation
with a source $F$.

%%%%%%%%%%%%%%%%%%%%%%%%%%%%%%%%%%%%%%%%%%%%%%%%%%%%%%%%%%%%%%%%%%%%%%%%%%%%
\section*{APPENDIX B. INTEGRAL TRANSFORMATION}

The Fourier transformation reads
$$
f(t,z)=\int\limits_{-\infty}^\infty\frac{d\mu d\mu^\prime}{(2\pi)^2}
\int\limits_{-\infty}^\infty f(p,s){\rm e}^{i\mu(t-p)-i\mu^\prime(z-s)}dpds.
$$

Let us introduce new variables:
\begin{eqnarray}
&&t=\tau\cosh{\theta},\quad z=\tau\sinh{\theta},\nonumber\\
&&p=\rho\cosh{\psi},\quad s=\rho\sinh{\psi},\nonumber\\
&&\mu=\lambda\cosh{\phi},\quad \mu^\prime=\lambda\sinh{\phi}.\nonumber
\end{eqnarray}

If $f(t,z)=F(\tau,\theta)$, one gets the following transformation rule:
\begin{eqnarray}
\tilde F(\lambda,\phi)=\int\limits_{-\infty}^\infty F(\rho,\psi)
{\rm e}^{-i\rho\lambda\cosh{(\phi-\psi)}}\rho d\rho d\psi,&&\nonumber\\
F(\tau,\theta)=\frac{1}{(2\pi)^2}\int\limits_{-\infty}^\infty\tilde F(\lambda,\phi)
{\rm e}^{i\tau\lambda\cosh{(\phi-\theta)}}\lambda d\lambda d\phi.&&\nonumber
\end{eqnarray}

For example, we find that
$$
\frac{1}{\tau}\delta(\Delta\tau)\delta(\Delta\theta)=\frac{1}{(2\pi)^2}\int\limits_{-\infty}^\infty
{\rm e}^{i\lambda[\tau\cosh{(\phi-\Delta\theta)}-\tau_0\cosh{\phi}]}\lambda d\lambda d\phi,
$$
where $\Delta\tau=\tau-\tau_0$, $\Delta\theta=\theta-\theta_0$.

%%%%%%%%%%%%%%%%%%%%%%%%%%%%%%%%%%%%%%%%%%%%%%%%%%%%%%%%%%%%%%%%%%%%%%%%%%%%

\end{document}